\documentclass[journal=apchd5,manuscript=letter,email=false,layout=twocolumn]{achemso}

\usepackage[version=3]{mhchem} 
\usepackage[T1]{fontenc}      
\usepackage{caption}
\usepackage{float}
\usepackage{geometry}
\usepackage{natbib}
\usepackage{setspace}
\usepackage{xkeyval}
\usepackage{graphicx}
\usepackage{epstopdf}
\usepackage[bookmarks=false,linkcolor=blue,urlcolor=blue,colorlinks,citecolor=blue]{hyperref}

\author{Michal Yachini}
\author{Boris Malomed}
\author{Alon Bahabad}
\affiliation{Department of Physical Electronics, School of Electrical Engineering,
	Fleischman Faculty of Engineering, Tel-Aviv University, Tel-Aviv 69978,
	Israel}

\title {Envelope time reversal of optical pulses following frequency conversion with accelerating quasi-phase-matching}

\begin{document}
	
\begin{abstract}
 It is shown theoretically that the use of accelerating spatiotemporal
 quasi-phase-matching (QPM) modulation patterns in media with parametric
 optical interactions makes it possible to generate a time-reversed replica
 of the pump pulse envelope in a frequency converted signal. The conversion is
 dependent on the group-velocity mismatch between the fundamental and up-converted
 harmonics, and controlled by the acceleration rate (\emph{chirp}) of the QPM phase
 pattern. Analytical results are corroborated by numerical simulations.
\end{abstract}

\section{}
Time reversal of pulses has important applications, such as the correction
of wave distortions \cite{agarwal1983scattering} and focusing in various
settings, including complex media \cite{aulbach2011control}, plasmonics \cite{li2008highly}, medical ultrasound \cite{fink1993time} and communications
microwaves \cite{lerosey2007focusing}. In optics, time reversal was
demonstrated or predicted by means of two different methods: by phase
conjugation via nonlinear four-wave mixing \cite{yariv1979compensation, miller1980time,kuzucu2009spectral,marom2000time,joubert1989temporal} or by
using time-modulated photonic structures \cite{yanik2004time,longhi2007stopping, sivan2011time,sanghoon2010true,zheng2013time,chumak2010all,yanik2005dynamic}. A recent related result is the inversion of Airy pulses in a linear optical fiber with third-order dispersion \cite{Radik}. Here we aim to show that the interaction of a pump pulse with an accelerating (chirped) spatiotemporal nonlinear photonic crystal \cite{spatiotemporal, accelerate} can generate a signal pulse which is an envelope-time-reversed \cite{sivan2011theory}, frequency-converted, replica of the pump pulse, provided that group-velocity mismatch is maintained between the pump and signal pulses. Our signal is the result of two actions - envelope time reversal and, considering  the bandwidth associated with the signal envelope, broadband frequency conversion. In passing we mention that regardless of time-reversal, there are to date a few known techniques for inducing broadband optical frequency conversion such as autoresonant or adiabatic frequency conversion \cite{yaakobi2010autoresonant,yaakobi2013complete,suchowski2014adiabatic,moses2012fully,rangelov2012broadband}. Our results are also relevant to cases where energy is exchanged between different modes  due to a dynamical modulation\cite{karenowska2012oscillatory,sivan2016nonlinear}.

Dispersion-induced phase mismatch inhibits efficient optical-frequency
conversion processes. To ameliorate the situation in energy-conserving
processes, one can use properly patterned spatial modulations of a parameter
relevant to the process to compensate for momentum mismatch. This technique
is known as Quasi-Phase-Matching (QPM) \cite{boyd,armstrong1962interactions}.%
More generally, the phase mismatch may be split between the momentum and
energy domains, in which case a spatiotemporal modulation is needed to
phase-match the process. Such spatiotemporal QPM was actually demonstrated
for high-harmonic-generation, prior to the full theoretical treatment \cite%
{spatiotemporal}, using a modulation in the form of a constant-velocity
grating realized by a train of counter-propagating pulses \cite%
{zhang2007quasi}. The availability of techniques for engineering complex
spatiotemporal light patterns \cite{akturk2010spatio,konsens2016time}
suggests that all-optical spatiotemporal QPM can be produced, using
modulations more sophisticated than gratings moving at a constant velocity.
In particular, an accelerating grating can enforce different phase-matching
(PM) conditions at different times in the course of the nonlinear
interaction. Spatiotemporal QPM with specific accelerating modulations were
suggested for controlling the temporal and spectral profiles of 
high-harmonic generation \cite{accelerate}, and for realizing
time-to-frequency mapping of optical pulses \cite{konsens2016time}.

In the following we show that, choosing an accelerating all-optical modulation pattern, one can realize a frequency converted signal having the time-reversed envelope of the optical pump
pulse interacting with the pattern. We stress that the method is relevant to
any frequency-conversion process in which the group-velocity-mismatch is
significant and an all-optical modulation is applicable, both in perturbative
\cite{bahabad2008quasi,Myer:14} and in extreme nonlinear optics \cite
{zhang2007quasi}. Without loss of generality we develop this concept for the prototypical nonlinear frequency conversion of Second-Harmonic-Generation (SHG).

We start with the one-dimensional wave equation for the Second-Harmonic (SH) field in the
frequency domain under the no-depletion approximation in a non-magnetic
medium:
\begin{subequations}
\label{wave}\notag
\begin{equation}
\frac{\partial ^{2}\tilde{E}_{2\omega _{0}}(z,\omega )}{\partial z^{2}}+\beta ^{2}(\omega )\tilde{E}_{2\omega_{0}}(z,\omega )=-\mu _{0}\omega ^{2}\tilde{P}_{\mathrm{NL}}(z,\omega ),  
\end{equation}
\end{subequations}%
where $n(\omega )$ is the index of refraction, $c=1/\sqrt{\mu
	_{0}\varepsilon _{0}}$ is the speed of light, $\mu _{0}$ is the vacuum
permeability, $\varepsilon _{0}$ is the vacuum permittivity, and $\beta (\omega
)=n(\omega )\omega /c$ is the wavenumber. $\tilde{P}_{\mathrm{NL}}(z,\omega )$
is the Fourier transform of the material second-order nonlinear
polarization. In the time domain it is defined as

\begin{subequations}
	\label{Pnl}\notag
\begin{equation}
{P}_{\mathrm{NL}}(z,t)=\varepsilon _{0}\chi ^{(2)}g(z,t)E_{\omega
	_{0}}^{2}(z,t),  
\end{equation}
\end{subequations}

where $E_{\omega _{0}}(z,t)$ is the fundamental-harmonic (FH) electric
field, $\chi ^{(2)}$ is the second-order electric susceptibility, and $%
g(z,t)=e^{i\Phi (z,t)}$ is the spatiotemporal modulation imposed by the QPM modulation 
onto the nonlinear polarization. The spatial and temporal frequencies of the
phase function $\Phi (z,t)$ can be used to phase match momentum and energy
components, respectively \cite{spatiotemporal}.

\begin{figure}[tbh]
	\centering
	\includegraphics[trim={46mm 87mm 50mm
		92mm},clip,width=0.5\textwidth]{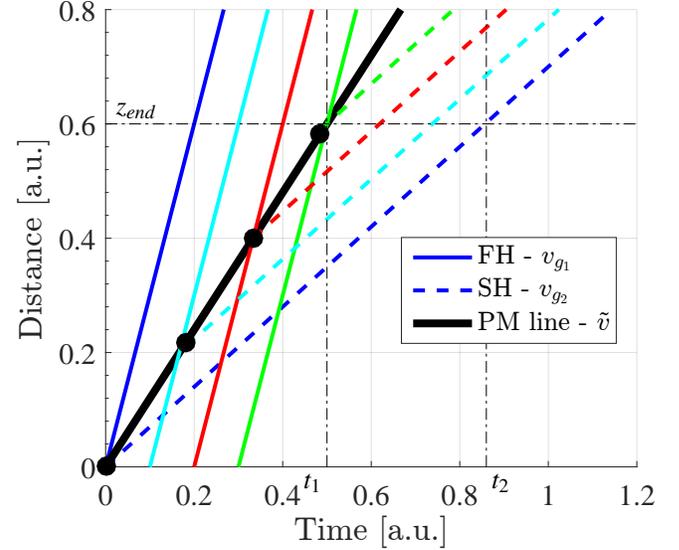}\llap{\parbox[c]{2.5cm}{%
			\vspace{-1.6cm}\footnotesize{$t_2$}}}\llap{\parbox[c]{5cm}{\vspace{-1.6cm}%
			\footnotesize{$t_1$}}}\llap{\parbox[c]{7.5cm}{\vspace{-11.5cm}%
			\footnotesize{$z_{end}$}}}
	\caption{Space-time diagram for envelope time reversal using an
		accelerating QPM modulation. Along the thick continuous line, corresponding
		to velocity $\tilde{v}$, the phase-matching (PM) condition is met for
		SHG. Continuous and dashed lines represent, respectively, wavelets (short segments of the emitted radiation) belonging
		to the FH pulse and SH radiation propagating at the respective group
		velocities. Whenever an FH wavelet encounters the PM line, an SH wavelet is
		emitted. At the end of the interaction, $z=z_{\mathrm{end}}$, the time ordering
		of the SH wavelets is the time reversal of the FH wavelets.}
	\label{WaveScheme}
\end{figure}

Our goal is to find a phase-modulation pattern, $\Phi (z,t)$, such that the
resulting SH temporal envelope will be the \emph{time reversal} of its
squared FH counterpart. First, we assume that, at $z=0$, the FH pulse starts
at $t=0$, and the FH (SH) moves at group velocity $v_{g_{1}}$ ($v_{g_{2}}$)
with $v_{g_{2}}<v_{g_{1}}$. We utilize the difference in group velocities
between the SH and the FH as follows: the QPM modulation, as we show below,
will satisfy PM conditions only for a short temporal interval around $t=z/%
\tilde{v}$ (the ``PM locus" denoted by the thick continuous
line in Fig.~\ref{WaveScheme}), where $v_{g_{2}}<\tilde{v}<v_{g_{1}}$. In
this small temporal interval, efficient up-conversion takes place.
Whenever a specific wavelet belonging to the FH pulse moving at velocity $%
v_{g{_{1}}}$ (continuous lines in Fig.~\ref{WaveScheme}) hits the PM line, an
SH wavelet is emitted, propagating at velocity $v_{g{_{2}}}$ (dashed lines in
Fig.~\ref{WaveScheme}). If the FH pulse width is $T_{1}$, the up-conversion
process will effectively cease at $z_{\mathrm{end}}=v_{g_{1}}\tilde{v}%
T_{1}/(v_{g_{1}}-\tilde{v})$. Using basic geometrical arguments it is apparent from Fig.~\ref{WaveScheme}
that, at this point, the SH envelope is the time reversal of the (squared) FH envelope, scaled with a factor
\begin{subequations}
	 \label{Widthfactor}\notag
\begin{equation}
R=\frac{\tilde{v}-v_{g_{2}}}{v_{g_{1}}-\tilde{v}}\cdot \frac{v_{g_{1}}}{%
	v_{g_{2}}}, 
\end{equation}%
\end{subequations}
such that the SH pulse duration is $T_{2}=RT_{1}$ (note that $t_{2}-t_{1}$
in Fig.~\ref{WaveScheme} is equal to $T_{2}$). Note that for $%
v_{g_{2}}>v_{g_{1}}$, the modulation needs to satisfy the condition $v_{g_{1}}<%
\tilde{v}<v_{g_{2}}$ while the PM line is $z=\tilde{v}(t-T_{1})$.
Now, we proceed to the identification of an appropriate modulation phase
function $\Phi (z,t)$ having the following spatial and temporal frequencies:

\begin{subequations}
	\label{derivativekw}
	\begin{align}
	\frac{\partial \Phi }{\partial z}& \equiv \Delta k(z,t), \\
	\frac{\partial \Phi }{\partial t}& \equiv -\Delta \omega (z,t).
	\end{align}%
\end{subequations}
	Meanwhile, if $\Delta k=2k(\omega _{0})-k({\tilde{\omega}})$ and $\Delta
	\omega =2\omega _{0}-\tilde{\omega}$ are the momentum and energy phase
	mismatches of the up-conversion process from frequency $\omega _{0}$
	to $\tilde{\omega}$, they must obey the phase-mismatch-compensation
	condition \cite{spatiotemporal}:

\begin{subequations}
	\label{PMcondition}\notag
	\begin{equation}
	\Delta k(z,t)=[\Delta \omega (z,t)n(\tilde{\omega})+2\omega _{0}\{n(\omega_{0})-n(\tilde{\omega})\}]/c.  	
	\end{equation}
\end{subequations}
As we require Eq. (\ref{PMcondition}) to hold solely along $z=\tilde{v}t$,
it is clear that the choice of
\begin{subequations}
	\label{modulation}\notag
\begin{equation}
\Phi (z,t)=\alpha (z-\tilde{v}t)^{2}+\Delta k_{0}z,  
\end{equation}%
\end{subequations}
with chirp constant (alias acceleration rate) $\alpha $, satisfies this
condition for $\Delta \omega =0$, provided that $\Delta k_{0}$ is the
momentum mismatch in the case of zero energy mismatch: $\Delta k_{0}=2\omega
_{0}[n(\omega _{0})-n(2\omega _{0})]/c $. In this case, the PM condition
brings the upconversion exactly to the second harmonic, $2\omega _{0}$.

Of course, this modulation format is not the only one possible for our
purpose, but it is, arguably, the simplest one. The temporal acceleration
rate of the modulation (chirp) is $\partial ^{2}\Phi /\partial t^{2}=2\alpha
\tilde{v}^{2}$. Faster acceleration moves points outside the PM line farther
from the PM conditions, making the accuracy of the envelope time reversal better. To
support a required temporal resolution $\Delta T$ associated with bandwidth $%
\Delta \Omega _{p}=2\pi /\Delta T$, the QPM-modulation bandwidth in this
interval, $\Delta \Omega _{m}=2\alpha \tilde{v}^{2}\Delta T$, must be much
larger: $\Delta \Omega _{m}\gg \Delta \Omega _{p}$. This resolution
condition may be quantified by a figure of merit, $F$:
\begin{subequations}
	\label{factor}\notag
\begin{equation}
F=\frac{\Delta \Omega _{m}}{\Delta \Omega _{p}}=\alpha \tilde{v}^{2}\Delta
T^{2}/\pi >>1  
\end{equation}%
\end{subequations}
Since $\tilde{v}$ is restricted by the material group velocities, for the
required resolution Eq. (\ref{factor}) imposes an essential condition on
the chirp constant $\alpha $ of the QPM modulation.

Apart from the desired phase-matched envelope time-reversal process, the proposed
modulation format may support other phase-matched upconversion processes.
This can be seen as Eqs.~(\ref{derivativekw}), (\ref{PMcondition}), and (\ref%
{modulation}) lead to the following condition:

\begin{subequations}
	\label{replicaCondition}\notag
\begin{equation}
2\omega _{0}\left[ \frac{1}{\tilde{v}}-\frac{n(2\omega _{0})}{c}\right] =%
\tilde{\omega}\left[ \frac{1}{\tilde{v}}-\frac{n(\tilde{\omega})}{c}\right] .
\end{equation}%
\end{subequations}
It is evident that the desired upconversion to $\tilde{\omega}=2\omega _{0}$
meets this criteria as planned. However, it is possible that the conversion
to other frequencies will satisfy this condition as well, depending on the
material dispersion and on the chosen velocity, $\tilde{v}$. Such
concomitant phase-matched processes will produce additional replicas of the
FH around different central frequencies. As long as these replicas stay well
separated in the frequency domain, the desired envelope-time-reversed signal can be
filtered out. The time orientation of any replica with respect to the FH
depends on the replica's group velocity $v_{g_{\tilde{\omega}}}$ (and its
relation to the FH group velocity and the spatiotemporal trajectory
determined by the PM condition).

A simple analytical model can explicitly demonstrate that the modulation
format proposed here indeed results in envelope time reversal. To this end, we
use the spatiotemporal slowly-varying-envelope approximation, along with
no-depletion approximation for the FH field. Also neglecting higher-order
dispersion, we reduce Eq.~(\ref{wave}) to:
\begin{subequations}
	 \label{first}\notag
\begin{equation}
\left( \frac{\partial }{\partial {z}}+\frac{1}{v_{g_{2}}}\frac{\partial }{%
	\partial t}\right) A_{2}(z,t)=\kappa e^{i\Phi (z,t)}A_{1}^{2}\left( t-\frac{z%
}{v_{g_{1}}}\right) , 
\end{equation}%
\end{subequations}
where $A_{1}$ ($A_{2}$) is the FH (SH) envelope, $\kappa \equiv -i\omega
_{SH}^{2}d_{\mathrm{eff}}/[c^{2}k(\omega _{\mathrm{SH}})]$ and $d_{\mathrm{eff}}
$ is the nonlinear-coupling coefficient. We make use of a coordinate
system moving with the SH group velocity, so that $\tau _{2}=t-z/v_{g_{2}}$%
, ~$\xi=z$. In this case,
Eq. (\ref{first}) becomes:
\begin{subequations}
	\label{SVEA_transformed}\notag
\begin{equation}
\frac{\partial }{\partial {\xi }}A_{2}(\xi ,\tau _{2})=\kappa e^{i\Phi (\xi
	,\tau _{2})}A_{1}^{2}\left[ \tau _{2}+\xi \left( \frac{1}{v_{g_{2}}}-\frac{1%
}{v_{g_{1}}}\right) \right] .  
\end{equation}%
\end{subequations}
Assuming that the acceleration of the QPM modulation pattern is large
enough, we use the stationary-phase approximation to integrate Eq. (\ref%
{SVEA_transformed}) around the stationary points $\xi _{s}$ determined by $%
\partial \Phi /\partial \xi |_{\xi _{s}}=0$, which yields

\begin{equation}
\xi _{s}=\frac{\Delta k_{0}}{2\alpha (1-\tilde{v}/v_{g_{2}})^{2}}+\frac{%
	\tilde{v}\tau _{2}}{1-\tilde{v}/v_{g_{2}}}\approx \frac{\tilde{v}\tau _{2}}{%
	1-\tilde{v}/v_{g_{2}}}
\end{equation}%

under the above condition, $\alpha \gg \Delta k_{0}$ (the stationary
point-equation is tantamount to the definition of the PM line, $z=\tilde{v}t$%
). Integration gives
\begin{subequations}
	 \label{solve}\notag
\begin{gather}
A_{2}(\xi ,\tau _{2})\approx A_{1}^{2}\left( \tau _{2}+\xi _{s}\left( \frac{1%
}{v_{g2}}-\frac{1}{v_{g1}}\right) \right) e^{-i\Phi (\xi _{s},\tau
_{2})}\kappa   \notag \\
\times \int_{0}^{\infty }\exp \left[ -\frac{i\Phi ^{\prime \prime }(\xi
	_{s},\tau _{2})}{2}(\xi -\xi _{s})^{2}\right] d\xi =  \notag \\
=\frac{\kappa }{2}\sqrt{\frac{2\pi }{|\Phi ^{\prime \prime }(\xi _{s},\tau
		_{2})|}}e^{\pi i/4}\exp \left[ i\Phi (\xi _{s},\tau _{2})\right]   \notag \\
\times A_{1}^{2}\left( \tau _{2}+\xi _{s}\left( \frac{1}{v_{g2}}-\frac{1}{%
	v_{g1}}\right) \right) . \tag{12}
\end{gather}%
\end{subequations}
Substituting $\xi _{s}$ and getting back to\ the $(z,t)$ coordinate system,
we obtain:

\begin{subequations}
	\label{solve 2}\notag
\begin{gather}
A_{2}(z,t)=\\
=\kappa \sqrt{\frac{\pi }{|\alpha |(1-\tilde{v}/v_{g_{2}})^{2}}}%
e^{\pi i/4}\exp \left[ i\Delta k_{0}\frac{\tilde{v}(v_{g_{2}}t-z)}{v_{g_{2}}-%
	\tilde{v}}\right]  \\
= A_{1}^{2}\left( -\frac{1}{R}\left( t-\frac{z}{v_{g_{2}}}\right) \right) ,
\tag{13}
\end{gather}
\end{subequations}
where the factor $R$, given by Eq.~(\ref{Widthfactor}), is positive for $%
v_{g_{2}}<\tilde{v}<v_{g_{1}}$, which secures the envelope time reversal. Having this result, a few remarks are in order. Essentially we are
interested in time reversal as concerns the absolute square of
the field, which is accomplished here. The absolute value of the
envelope of the SH is not specially sensitive to possible
sign changes in the FH (zero crossings). As concerns the phase profile
of the SH envelope, it is twice the time-reversed phase of the FH envelope, with an added
linear term. We also note that chirp added to the FH field will modify
the actual PM condition, which will then deviate from a straight line in
space-time, resulting in some distortion in the envelope-time-reversed wave
form. Still, for large enough constant $\alpha$ of the
accelerating modulation (which represents the intrinsic chirp of the
transformation) such distortions will be negligible.
Finally, we observe that the SH amplitude scales as $1/|\alpha|$, hence
better resolution due to larger $\alpha$ comes at the
price of a lower amplitude of the generated SH.

To demonstrate envelope time reversal using our proposed accelerating modulation
format, we have performed direct numerical integration of the full wave
equation (\ref{wave}), using the procedure outlined in Ref. \cite%
{konsens2016time}. As an input, we took an FH pulse with central
wavelength at $800$ nm, propagating in Barium borate (BBO) \cite%
{eimerl1987optical}. The modulation format is introduced with the help of
the phase function $\Phi $ in Eq.~(\ref{Pnl}), while in other settings
the amplitude modulation of the nonlinear polarization may also be used \cite{bahabad2008quasi,Myer:14}. 

\begin{figure}[tbh]
	\centering
	\includegraphics[trim={46mm 45mm 55mm
		45mm},clip,width=0.4\textwidth]{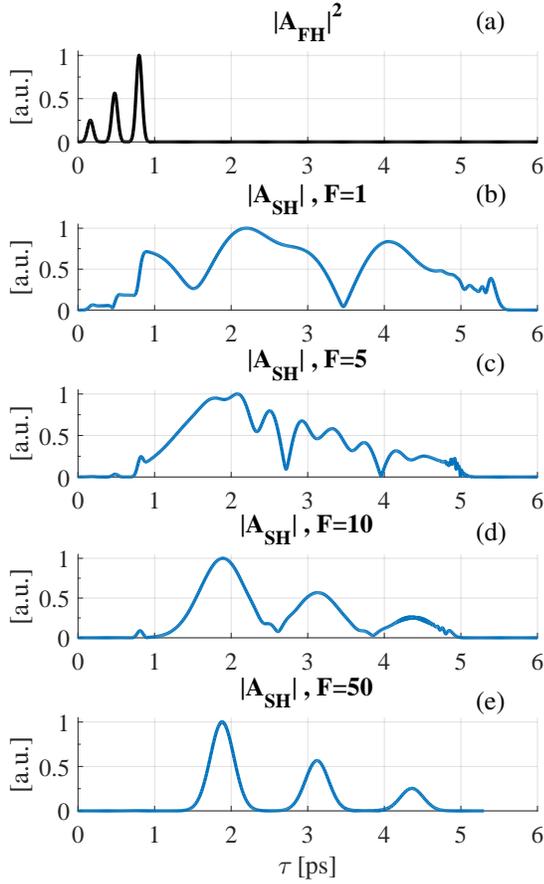} \llap{\parbox[c]{1cm}{%
			\vspace{-22.6cm}\footnotesize{(a)}}}\llap{\parbox[c]{1cm}{\vspace{-18cm}%
			\footnotesize{(b)}}}\llap{\parbox[c]{1cm}{\vspace{-13.5cm}%
			\footnotesize{(c)}}}\llap{\parbox[c]{1cm}{\vspace{-9cm}\footnotesize{(d)}}}%
	\llap{\parbox[c]{1cm}{\vspace{-4.5cm}\footnotesize{(e)}}}
	\caption{Envelope time reversal for varying acceleration rates of the QPM
		modulation. (a) The squared absolute value of the field in the FH pulse at
		the start of the interaction. (b)-(e) The absolute value of the SH field
		after completing the interaction with the accelerating QPM
		modulation structure. The results are characterized by values of the
		resolution-figure-of-merit, $F$, at different acceleration rates.}
	\label{FactorFig}
\end{figure}

First we look at the FH pulse of an overall duration $\sim 1$ ps with an
asymmetric envelope containing three peaks, distanced $0.3$ ps apart, with
increasing amplitudes, see Fig.~\hyperref[FactorFig]{\ref{FactorFig}(a)}.
The results are displayed in the reference frame moving at the FH group
velocity, so that $\tau =t-z/v_{g_{1}}$, $\xi =z$. The group velocities are 
$v_{g_{1}}=1.78\cdot 10^{8}$ 
m/s and $v_{g_{2}}=1.68\cdot 10^{8}$ m/s. 
The value of $\tilde{v}=1.76\cdot 10^{8}$ m/s, substituted in
Eq. (\ref{Widthfactor}), yields the scaling factor $R\approx 4$. We used four
accelerating QPM modulation formats with increasing values of the chirp rate $\alpha $%
, so that the corresponding resolution figure of merit $F$, defined in Eq.~(%
\ref{factor}), increases from $1$ to $50$, keeping the target resolution of $%
\Delta T=0.3$ ps. The respective shapes of the SH envelope produced by the
interaction are displayed in Fig.~\hyperref[FactorFig]{\ref{FactorFig}(b-e)}%
. It is evident that a large enough factor, $F=50$, secures obtaining an
exact envelope-time-reversed replica of the squared FH envelope. To estimate typical
parameters, we notice that, in the case of the FH pulse of duration $\sim 1$
ps, with $\Delta T=0.3$ ps and $F=10$, for which the envelope-time-reversed replica
has decent resolution, the overall bandwidth of the QPM modulation is $210$
nm for the central wavelength of $800$ nm. Such bandwidths are readily
achievable with commercial femtosecond lasers.

\begin{figure}[!hbt]
	\begin{tabular}{cccc}
		\includegraphics[trim={54mm 97mm 60mm 98mm},clip,width=0.16\textwidth]{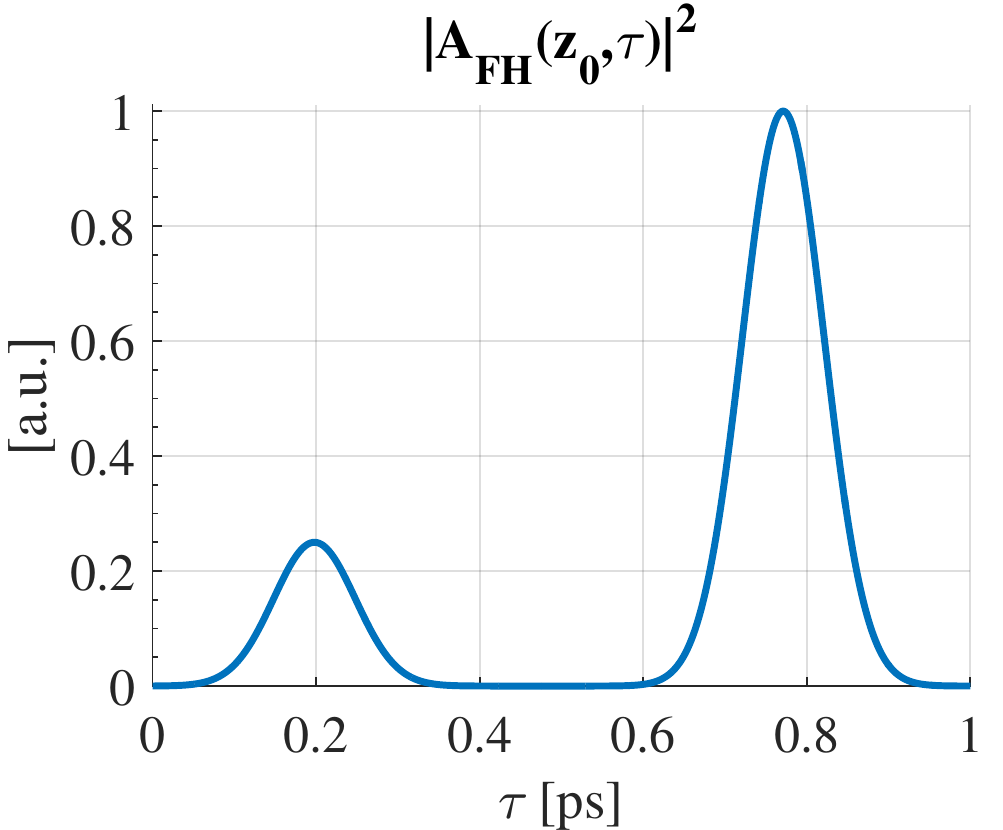}&
		\includegraphics[trim={54mm 97mm 60mm 98mm},clip,width=0.16\textwidth]{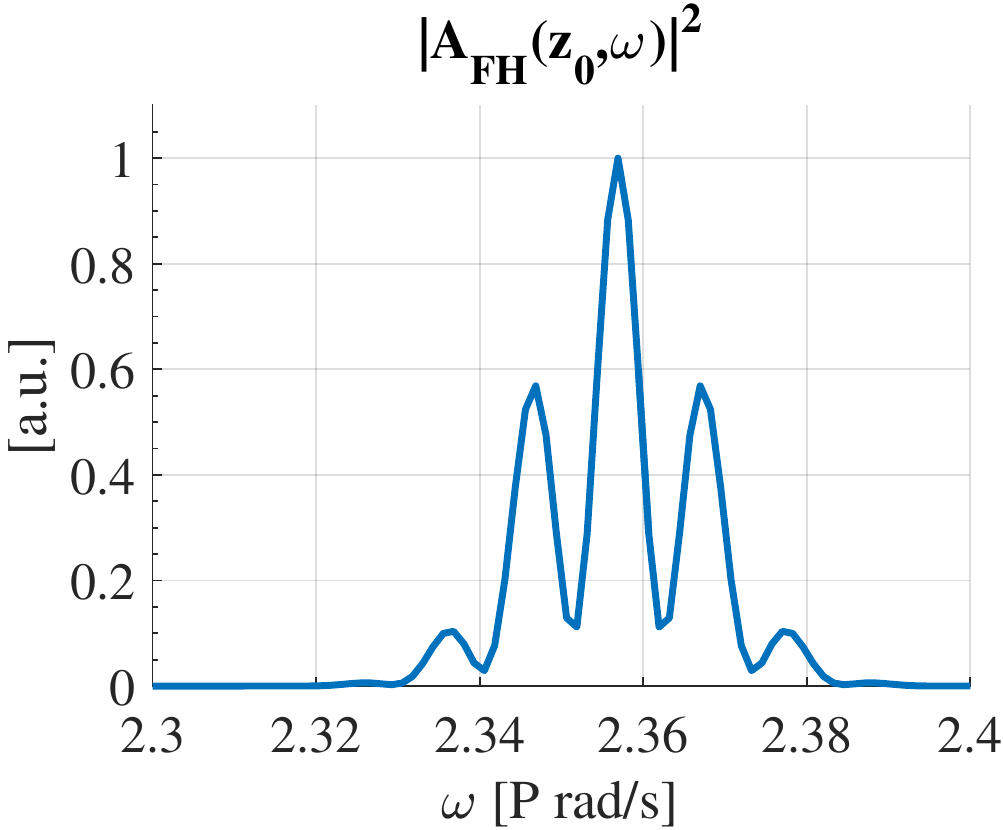}\\ \hline\\
		\includegraphics[trim={54mm 97mm 60mm 97mm},clip,width=0.16\textwidth]{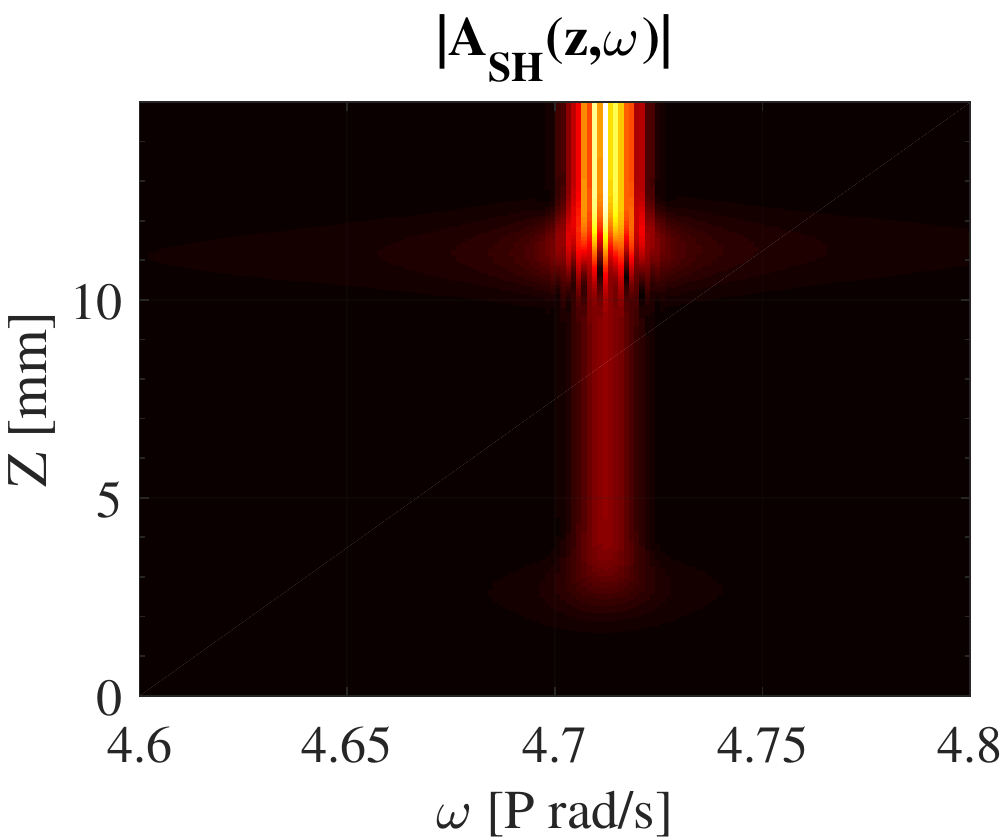}&
		\includegraphics[trim={54mm 97mm 60mm 97mm},clip,width=0.16\textwidth]{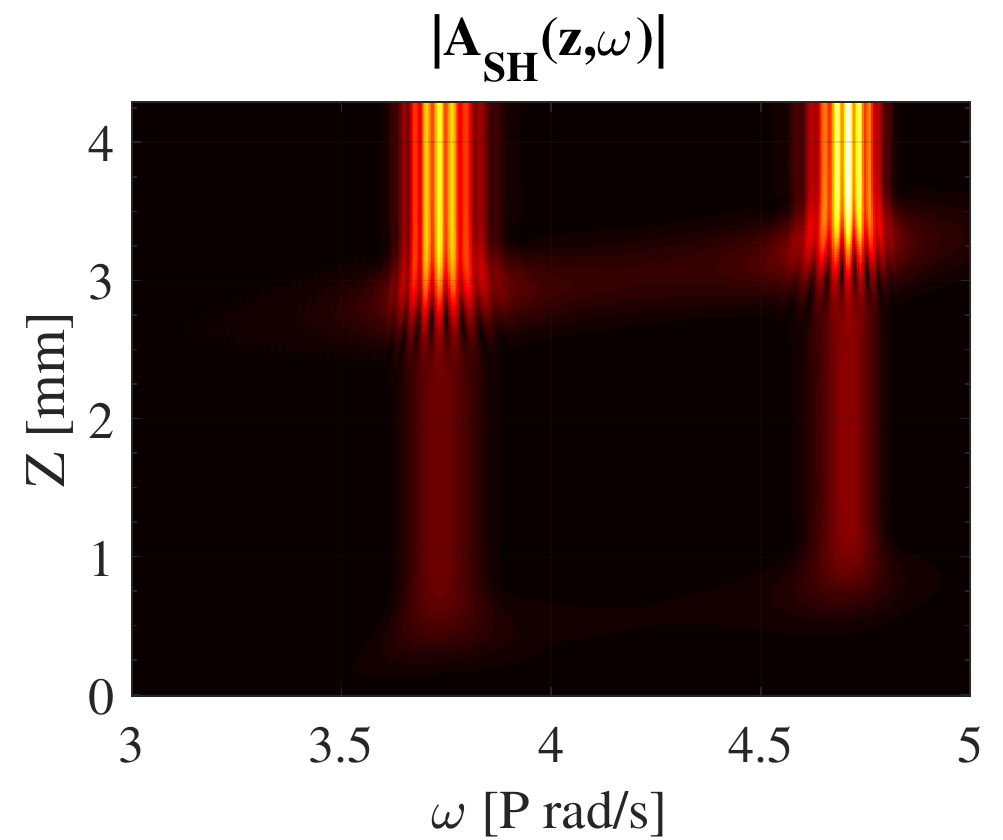}\\ 
		\includegraphics[trim={54mm 97mm 60mm 98mm},clip,width=0.16\textwidth]{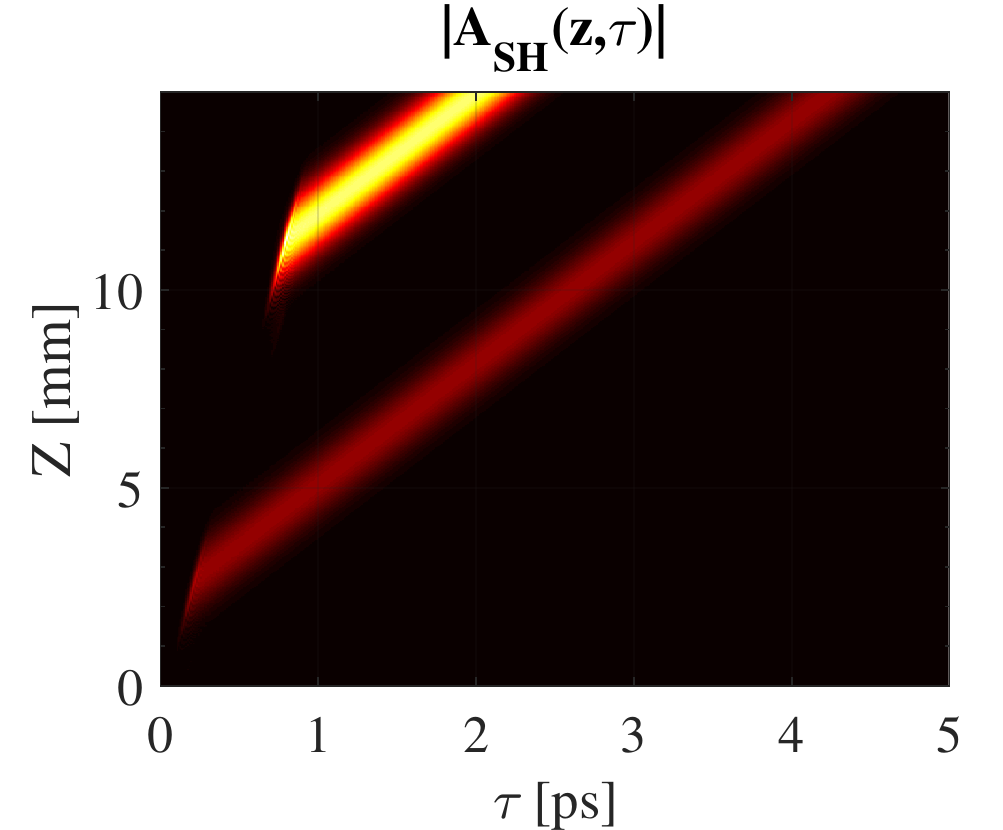}&
		\includegraphics[trim={54mm 97mm 60mm 98mm},clip,width=0.16\textwidth]{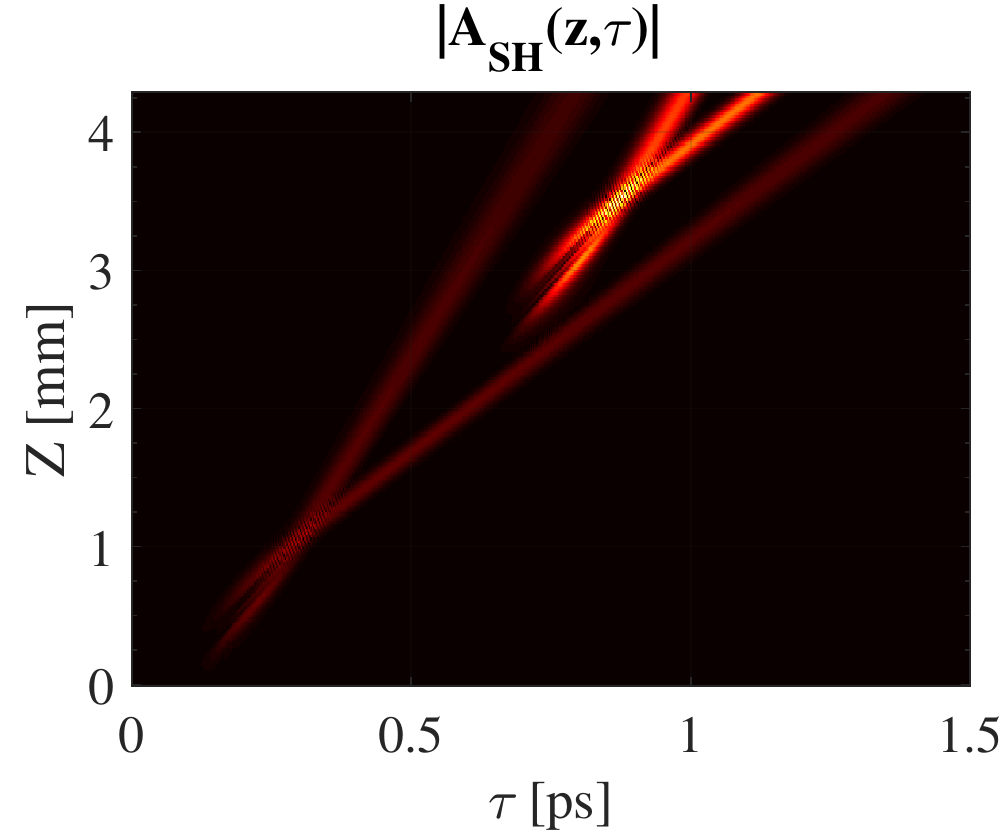}\\
		\includegraphics[trim={54mm 97mm 60mm 98mm},clip,width=0.16\textwidth]{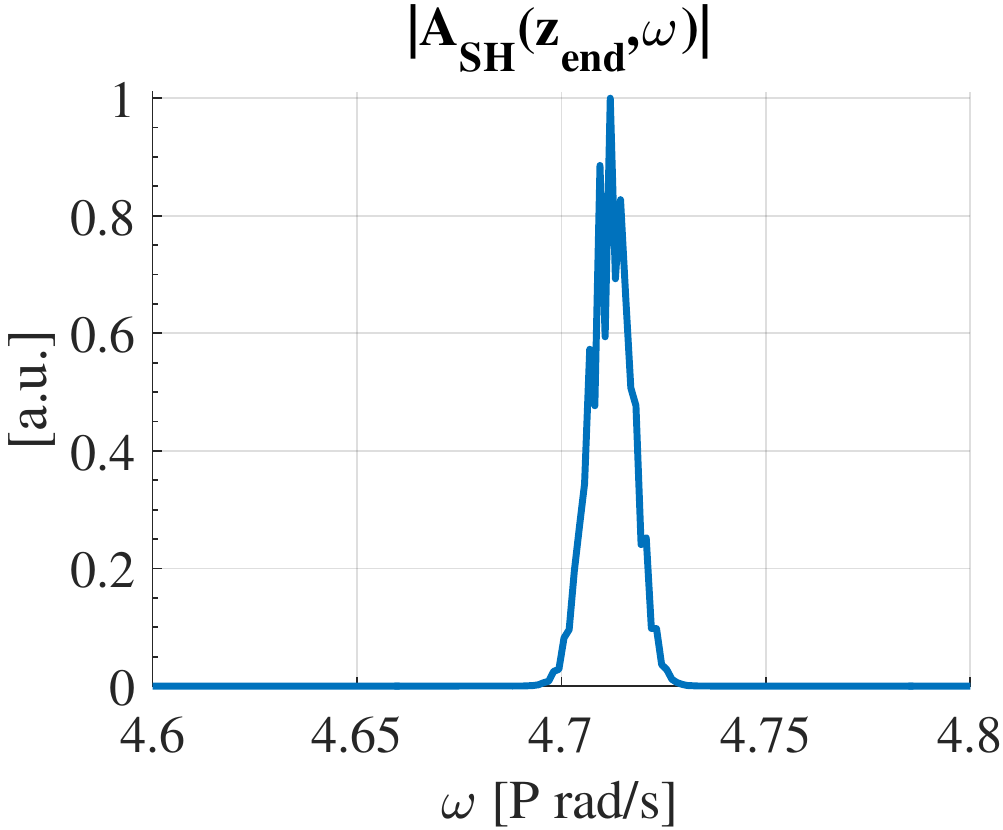}&
		\includegraphics[trim={54mm 97mm 60mm 98mm},clip,width=0.16\textwidth]{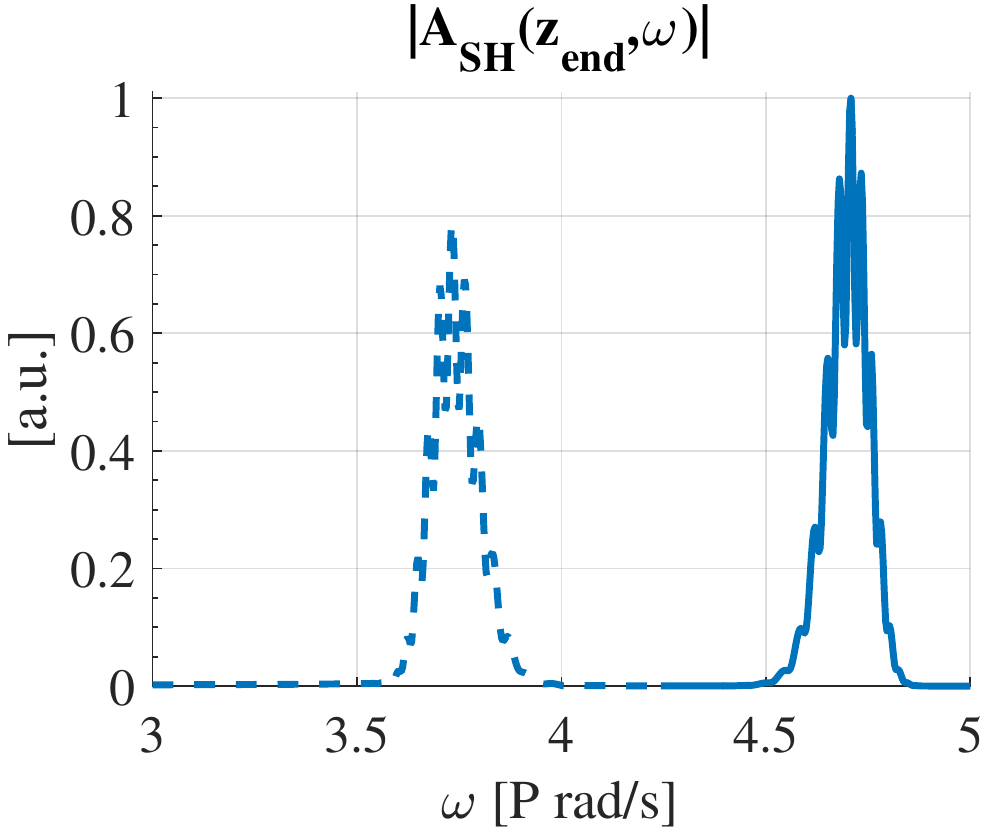}\\
		\includegraphics[trim={54mm 97mm 60mm 98mm},clip,width=0.16\textwidth]{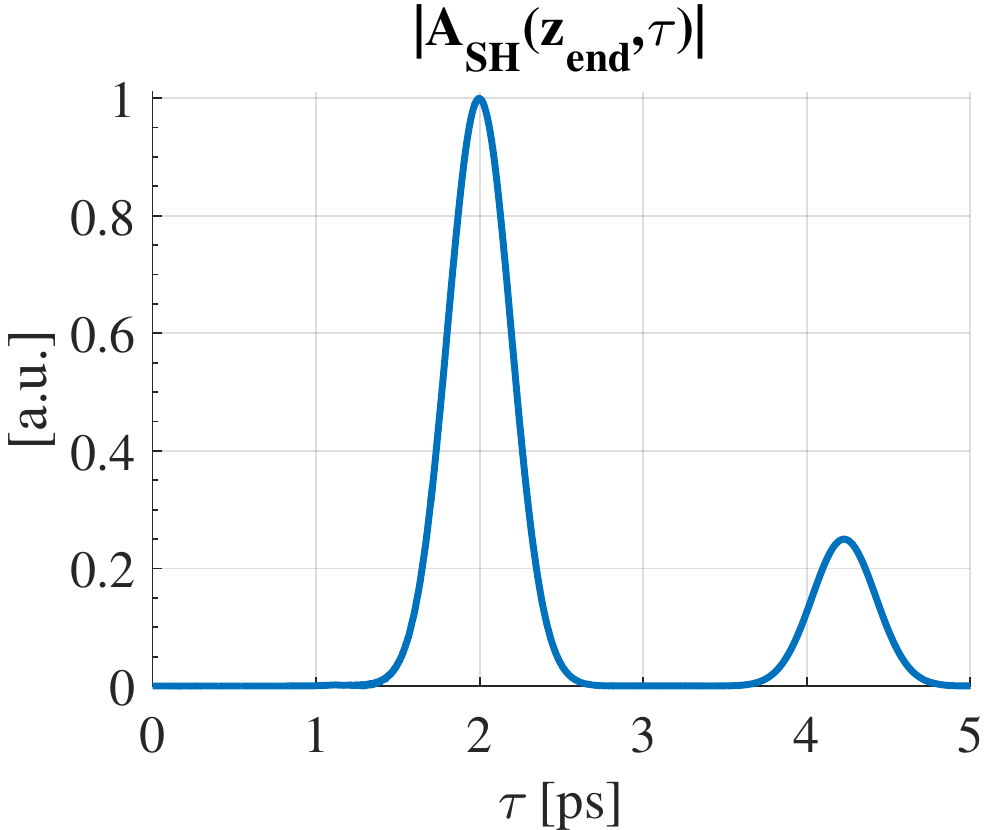}&
		\includegraphics[trim={54mm 97mm 60mm 98mm},clip,width=0.16\textwidth]{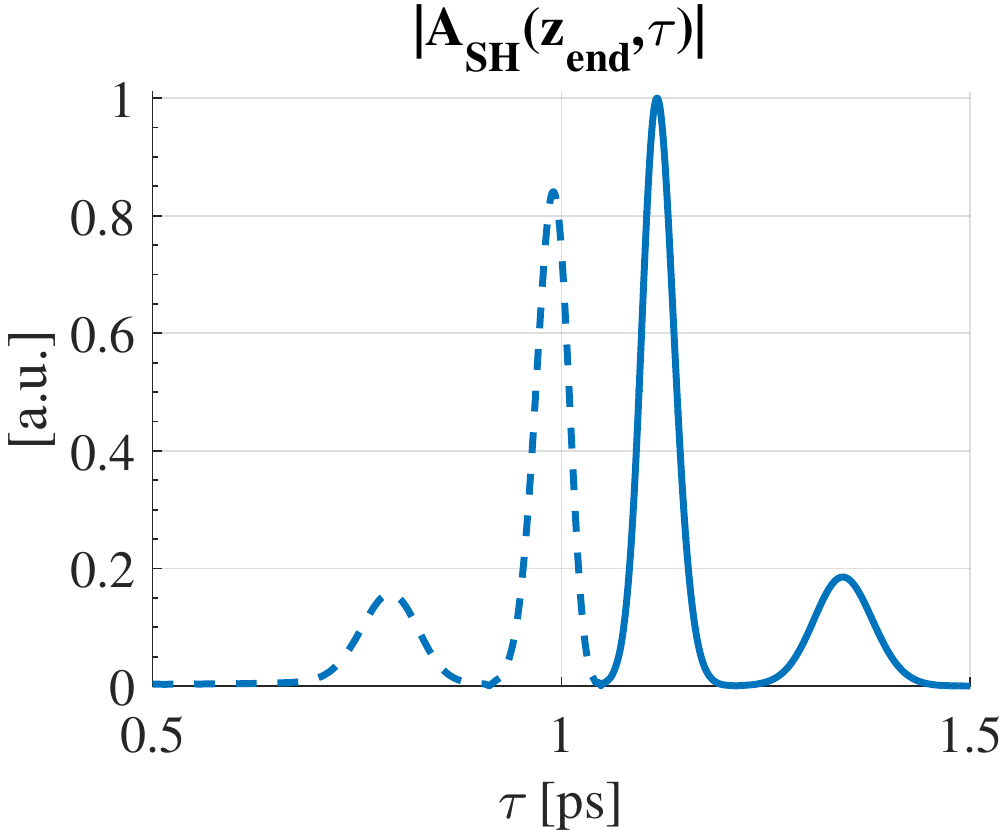}
	\end{tabular}
	\llap{\parbox[c]{6cm}{\vspace{-13cm}\footnotesize{(a)}}}\llap{\parbox[c]{3cm}{\vspace{-13cm}\footnotesize{(b)}}}\llap{\parbox[c]{6cm}{\vspace{-7cm}\footnotesize{(c)}}}\llap{\parbox[c]{3cm}{\vspace{-7cm}\footnotesize{(d)}}}\llap{\parbox[c]{6cm}{\vspace{-2cm}\footnotesize{(e)}}}\llap{\parbox[c]{3cm}{\vspace{-2cm}\footnotesize{(f)}}}\llap{\parbox[c]{6cm}{\vspace{3cm}\footnotesize{(g)}}}\llap{\parbox[c]{3cm}{\vspace{3cm}\footnotesize{(h)}}}\llap{\parbox[c]{6cm}{\vspace{8cm}\footnotesize{(i)}}}\llap{\parbox[c]{3cm}{\vspace{8cm}\footnotesize{(j)}}}
	\caption{Full temporal and spectral evolution of radiation emitted due to
		the interaction of an FH with the accelerating QPM modulation. Top
		panel: (a) temporal and (b) spectral dependence of the FH field. Bottom
		panel: left column: evolution of the SH for a QPM modulation allowing for a
		single envelope-time-reversed SH replica of the FH; right column: evolution of the
		up-converted radiation for the case of a QPM modulation allowing for an
		additional non-reversed replica of the FH. (c),(d) spectral evolution
		(e),(f) temporal evolution (g),(h) the up-converted spectrum at the end of
		the interaction (i),(j) the up-converted amplitude in the time domain at the
		end of the interaction. (The non-reversed replica is marked with a dashed
		line) }
	\label{ReplicaFig}
\end{figure}

Next we will look at the full temporal and spectral evolution of the SH
field along the interaction coordinate for two different velocities, $\tilde{%
	v}=1.76\cdot 10^{8}$ and $1.71\cdot 10^{8}$ m/s, while $\alpha =5\cdot
10^{10}~ $rad/m$^{2}$ for both cases. Together with the target value of $%
\Delta T=0.55 $ ps, one has the corresponding values $F=150$ and $140$ for
the two cases. This time, the FH field has an asymmetric double-peak
envelope with a $1 $ ps  duration, see Fig.~\hyperref[ReplicaFig]{\ref{ReplicaFig}%
	(a)}. For $\tilde{v}=1.76\cdot 10^{8}$ m/s the QPM modulation gives rise
precisely to the intended envelope-time-reversed shape of the SH, as seen in Fig.~\hyperref[ReplicaFig]{\ref{ReplicaFig}(i)}. In this case, the temporal
evolution clearly shows that the two peaks of the FH profile are generated at
different coordinates. Although the earlier FH peak is upconverted first, at
the end of the interaction it lags behind the upconverted second peak,
because the SH group velocity is smaller than both the FH group velocity and
the velocity $\tilde{v}$ selected by the PM condition (\ref{PMcondition}). Thus,
the sequence of the two peaks in the SH envelope is reversed versus the
original FH envelope.

For the second case, with $\tilde{v}=1.71\cdot 10^{8}$ m/s%
, the modulation supports the envelope-time-reversed replica at $\tilde{\omega}%
=2\omega _{0}$, shown by a continuous line in Fig.~\hyperref[ReplicaFig]{\ref%
	{ReplicaFig}(j)}, and an additional spectrally separated, non-reversed
replica, shown by a dashed line in Fig.~\hyperref[ReplicaFig]{\ref%
	{ReplicaFig}(j)}, at $\tilde{\omega}=1.6\omega _{0}$, in accordance with
Eq.~(\ref{replicaCondition}). The second replica is not inverted because its
group velocity, $1.735\cdot 10^{8} $ m/s, is higher than the PM condition velocity $\tilde{v}$ although the order of generation of the two peaks in the SH is the same as in the inverted replica. 
We note that the QPM modulations
are different for the two cases, as they depend on the value of $%
\tilde{v}$, leading also to a difference in the scaling factors: $R\approx
4\,(0.4)$ for $\tilde{v}=1.76\cdot 10^{8}\,(1.71\cdot 10^{8})$ m/s.

As noted above, our model neglects high-order dispersion terms. This means that for a reliable envelope-time-reversal the interaction length should be shorter than the dispersion length $l_D \equiv T_1^2/\beta_2$ ($T_1$ is the FH pulse duration and $\beta_2$ is its group velocity dispersion) at which dispersion starts to significantly distorts the pump pulse. In typical situations such as considered here (i.e. picosecond duration pump pulse propagating in BBO), this is far from being an actual limitation as the interaction length is about a centimeter  while the dispersion length is several meters. 

To conclude, we have shown that accelerating (chirped) spatiotemporal
quasi-phase-matching modulation can be used for up-converting an FH field to
an envelope-time-reversed replica under the non-depletion approximation and when higher-order dispersion is negligible. The choice of
the appropriate QPM modulation is dependent on the group-velocity
mismatch between the interacting fields. Similarly to other proposed
methods of time reversal \cite{yariv1979compensation,
	miller1980time,kuzucu2009spectral,yanik2004time, longhi2007stopping,
	sivan2011time}, the proposed modulation needs to be synchronized with the
time of arrival of the FH field, and its bandwidth must be much larger than
the bandwidth of the FH, to secure high-quality inversion. The latter
condition is controlled by the chirp rate of the QPM modulation. While
we have demonstrated the proposed scheme in detail using the most
fundamental nonlinear process of SHG, it should be relevant to any 
nonlinear optical process based on parametric interactions, such as 
high-harmonic generation, where the nonlinear polarization can be manipulated macroscopically using an 
all-optical perturbation with high efficiency \cite{zhang2007quasi}. Finally, we note that the use of all-optical accelerating modulations that we discussed here might also find use in other scenarios including chromatic dispersion compensation  \cite{conjugation, kuzucu2009spectral}, and solitons manipulation \cite{Chu}.

M.Y and A.B acknowledge support from the Israel Science Foundation (ISF) (1233/13).
The work of B.A.M. is supported, in part, by grant No. 2015616 from the joint program in physics 
between NSF and Binational (US-Israel) Science Foundation.

\providecommand{\latin}[1]{#1}
\makeatletter
\providecommand{\doi}
{\begingroup\let\do\@makeother\dospecials
	\catcode`\{=1 \catcode`\}=2\doi@aux}
\providecommand{\doi@aux}[1]{\endgroup\texttt{#1}}
\makeatother
\providecommand*\mcitethebibliography{\thebibliography}
\csname @ifundefined\endcsname{endmcitethebibliography}
{\let\endmcitethebibliography\endthebibliography}{}


\begin{mcitethebibliography}{39}
	\providecommand*\natexlab[1]{#1}
	\providecommand*\mciteSetBstSublistMode[1]{}
	\providecommand*\mciteSetBstMaxWidthForm[2]{}
	\providecommand*\mciteBstWouldAddEndPuncttrue
	{\def\EndOfBibitem{\unskip.}}
	\providecommand*\mciteBstWouldAddEndPunctfalse
	{\let\EndOfBibitem\relax}
	\providecommand*\mciteSetBstMidEndSepPunct[3]{}
	\providecommand*\mciteSetBstSublistLabelBeginEnd[3]{}
	\providecommand*\EndOfBibitem{}
	\mciteSetBstSublistMode{f}
	\mciteSetBstMaxWidthForm{subitem}{(\alph{mcitesubitemcount})}
	\mciteSetBstSublistLabelBeginEnd
	{\mcitemaxwidthsubitemform\space}
	{\relax}
	{\relax}
	
	\bibitem[Agarwal \latin{et~al.}(1983)Agarwal, Friberg, and
	Wolf]{agarwal1983scattering}
	Agarwal,~G.; Friberg,~A.~T.; Wolf,~E. Scattering theory of distortion
	correction by phase conjugation. \emph{J. Opt. Soc. Am.} \textbf{1983},
	\emph{73}, 529--538\relax
	\mciteBstWouldAddEndPuncttrue
	\mciteSetBstMidEndSepPunct{\mcitedefaultmidpunct}
	{\mcitedefaultendpunct}{\mcitedefaultseppunct}\relax
	\EndOfBibitem
	\bibitem[Aulbach \latin{et~al.}(2011)Aulbach, Gjonaj, Johnson, Mosk, and
	Lagendijk]{aulbach2011control}
	Aulbach,~J.; Gjonaj,~B.; Johnson,~P.~M.; Mosk,~A.~P.; Lagendijk,~A. Control of
	light transmission through opaque scattering media in space and time.
	\emph{Phys. Rev. Lett.} \textbf{2011}, \emph{106}, 103901\relax
	\mciteBstWouldAddEndPuncttrue
	\mciteSetBstMidEndSepPunct{\mcitedefaultmidpunct}
	{\mcitedefaultendpunct}{\mcitedefaultseppunct}\relax
	\EndOfBibitem
	\bibitem[Li and Stockman(2008)Li, and Stockman]{li2008highly}
	Li,~X.; Stockman,~M.~I. Highly efficient spatiotemporal coherent control in
	nanoplasmonics on a nanometer-femtosecond scale by time reversal. \emph{Phys.
		Rev. B} \textbf{2008}, \emph{77}, 195109\relax
	\mciteBstWouldAddEndPuncttrue
	\mciteSetBstMidEndSepPunct{\mcitedefaultmidpunct}
	{\mcitedefaultendpunct}{\mcitedefaultseppunct}\relax
	\EndOfBibitem
	\bibitem[Fink(1993)]{fink1993time}
	Fink,~M. Time-reversal mirrors. \emph{J. Phys. D: Appl. Phys.} \textbf{1993},
	\emph{26}, 1333\relax
	\mciteBstWouldAddEndPuncttrue
	\mciteSetBstMidEndSepPunct{\mcitedefaultmidpunct}
	{\mcitedefaultendpunct}{\mcitedefaultseppunct}\relax
	\EndOfBibitem
	\bibitem[Lerosey \latin{et~al.}(2007)Lerosey, De~Rosny, Tourin, and
	Fink]{lerosey2007focusing}
	Lerosey,~G.; De~Rosny,~J.; Tourin,~A.; Fink,~M. Focusing beyond the diffraction
	limit with far-field time reversal. \emph{Science} \textbf{2007}, \emph{315},
	1120--1122\relax
	\mciteBstWouldAddEndPuncttrue
	\mciteSetBstMidEndSepPunct{\mcitedefaultmidpunct}
	{\mcitedefaultendpunct}{\mcitedefaultseppunct}\relax
	\EndOfBibitem
	\bibitem[Yariv \latin{et~al.}(1979)Yariv, Fekete, and
	Pepper]{yariv1979compensation}
	Yariv,~A.; Fekete,~D.; Pepper,~D.~M. Compensation for channel dispersion by
	nonlinear optical phase conjugation. \emph{Opt. Lett.} \textbf{1979},
	\emph{4}, 52--54\relax
	\mciteBstWouldAddEndPuncttrue
	\mciteSetBstMidEndSepPunct{\mcitedefaultmidpunct}
	{\mcitedefaultendpunct}{\mcitedefaultseppunct}\relax
	\EndOfBibitem
	\bibitem[Miller(1980)]{miller1980time}
	Miller,~D. Time reversal of optical pulses by four-wave mixing. \emph{Opt.
		Lett.} \textbf{1980}, \emph{5}, 300--302\relax
	\mciteBstWouldAddEndPuncttrue
	\mciteSetBstMidEndSepPunct{\mcitedefaultmidpunct}
	{\mcitedefaultendpunct}{\mcitedefaultseppunct}\relax
	\EndOfBibitem
	\bibitem[Kuzucu \latin{et~al.}(2009)Kuzucu, Okawachi, Salem, Foster,
	Turner-Foster, Lipson, and Gaeta]{kuzucu2009spectral}
	Kuzucu,~O.; Okawachi,~Y.; Salem,~R.; Foster,~M.~A.; Turner-Foster,~A.~C.;
	Lipson,~M.; Gaeta,~A.~L. Spectral phase conjugation via temporal imaging.
	\emph{Opt. Express} \textbf{2009}, \emph{17}, 20605--20614\relax
	\mciteBstWouldAddEndPuncttrue
	\mciteSetBstMidEndSepPunct{\mcitedefaultmidpunct}
	{\mcitedefaultendpunct}{\mcitedefaultseppunct}\relax
	\EndOfBibitem
	\bibitem[Marom \latin{et~al.}(2000)Marom, Panasenko, Rokitski, Sun, and
	Fainman]{marom2000time}
	Marom,~D.; Panasenko,~D.; Rokitski,~R.; Sun,~P.-C.; Fainman,~Y. Time reversal
	of ultrafast waveforms by wave mixing of spectrally decomposed waves.
	\emph{Opt. Lett.} \textbf{2000}, \emph{25}, 132--134\relax
	\mciteBstWouldAddEndPuncttrue
	\mciteSetBstMidEndSepPunct{\mcitedefaultmidpunct}
	{\mcitedefaultendpunct}{\mcitedefaultseppunct}\relax
	\EndOfBibitem
	\bibitem[Joubert \latin{et~al.}(1989)Joubert, Roblin, and
	Grousson]{joubert1989temporal}
	Joubert,~C.; Roblin,~M.~L.; Grousson,~R. Temporal reversal of picosecond
	optical pulses by holographic phase conjugation. \emph{Appl. Opt.}
	\textbf{1989}, \emph{28}, 4604--4612\relax
	\mciteBstWouldAddEndPuncttrue
	\mciteSetBstMidEndSepPunct{\mcitedefaultmidpunct}
	{\mcitedefaultendpunct}{\mcitedefaultseppunct}\relax
	\EndOfBibitem
	\bibitem[Yanik and Fan(2004)Yanik, and Fan]{yanik2004time}
	Yanik,~M.~F.; Fan,~S. Time reversal of light with linear optics and modulators.
	\emph{Phys. Rev. Lett.} \textbf{2004}, \emph{93}, 173903\relax
	\mciteBstWouldAddEndPuncttrue
	\mciteSetBstMidEndSepPunct{\mcitedefaultmidpunct}
	{\mcitedefaultendpunct}{\mcitedefaultseppunct}\relax
	\EndOfBibitem
	\bibitem[Longhi(2007)]{longhi2007stopping}
	Longhi,~S. Stopping and time reversal of light in dynamic photonic structures
	via Bloch oscillations. \emph{Phys. Rev. E} \textbf{2007}, \emph{75},
	026606\relax
	\mciteBstWouldAddEndPuncttrue
	\mciteSetBstMidEndSepPunct{\mcitedefaultmidpunct}
	{\mcitedefaultendpunct}{\mcitedefaultseppunct}\relax
	\EndOfBibitem
	\bibitem[Sivan and Pendry(2011)Sivan, and Pendry]{sivan2011time}
	Sivan,~Y.; Pendry,~J.~B. Time reversal in dynamically tuned zero-gap periodic
	systems. \emph{Phys. Rev. Lett.} \textbf{2011}, \emph{106}, 193902\relax
	\mciteBstWouldAddEndPuncttrue
	\mciteSetBstMidEndSepPunct{\mcitedefaultmidpunct}
	{\mcitedefaultendpunct}{\mcitedefaultseppunct}\relax
	\EndOfBibitem
	\bibitem[Sanghoon \latin{et~al.}(2010)Sanghoon, Primerov, Song, Th{\'e}venaz,
	Santagiustina, and Ursini]{sanghoon2010true}
	Sanghoon,~C.; Primerov,~N.; Song,~K.~Y.; Th{\'e}venaz,~L.; Santagiustina,~M.;
	Ursini,~L. True time reversal via dynamic Brillouin gratings in polarization
	maintaining fibers. Nonlinear Photonics. 2010; p NThA6\relax
	\mciteBstWouldAddEndPuncttrue
	\mciteSetBstMidEndSepPunct{\mcitedefaultmidpunct}
	{\mcitedefaultendpunct}{\mcitedefaultseppunct}\relax
	\EndOfBibitem
	\bibitem[Zheng \latin{et~al.}(2013)Zheng, Ren, Wan, and Chen]{zheng2013time}
	Zheng,~Y.; Ren,~H.; Wan,~W.; Chen,~X. Time-reversed wave mixing in nonlinear
	optics. \emph{Sci. Rep.} \textbf{2013}, \emph{3}\relax
	\mciteBstWouldAddEndPuncttrue
	\mciteSetBstMidEndSepPunct{\mcitedefaultmidpunct}
	{\mcitedefaultendpunct}{\mcitedefaultseppunct}\relax
	\EndOfBibitem
	\bibitem[Chumak \latin{et~al.}(2010)Chumak, Tiberkevich, Karenowska, Serga,
	Gregg, Slavin, and Hillebrands]{chumak2010all}
	Chumak,~A.~V.; Tiberkevich,~V.~S.; Karenowska,~A.~D.; Serga,~A.~A.;
	Gregg,~J.~F.; Slavin,~A.~N.; Hillebrands,~B. All-linear time reversal by a
	dynamic artificial crystal. \emph{Nat. Commun.} \textbf{2010}, \emph{1},
	141\relax
	\mciteBstWouldAddEndPuncttrue
	\mciteSetBstMidEndSepPunct{\mcitedefaultmidpunct}
	{\mcitedefaultendpunct}{\mcitedefaultseppunct}\relax
	\EndOfBibitem
	\bibitem[Yanik and Fan(2005)Yanik, and Fan]{yanik2005dynamic}
	Yanik,~M.~F.; Fan,~S. Dynamic photonic structures: stopping, storage, and time
	reversal of light. \emph{Stud. Appl. Math.} \textbf{2005}, \emph{115},
	233--253\relax
	\mciteBstWouldAddEndPuncttrue
	\mciteSetBstMidEndSepPunct{\mcitedefaultmidpunct}
	{\mcitedefaultendpunct}{\mcitedefaultseppunct}\relax
	\EndOfBibitem
	\bibitem[Driben \latin{et~al.}(2012)Driben, Hu, Chen, Malomed, and
	Morandotti]{Radik}
	Driben,~R.; Hu,~Y.; Chen,~Z.; Malomed,~B.~A.; Morandotti,~R. Inversion and
	tight focusing of Airy pulses under the action of third-order dispersion.
	\emph{Opt. Lett.} \textbf{2012}, \emph{38}, 2499--2501\relax
	\mciteBstWouldAddEndPuncttrue
	\mciteSetBstMidEndSepPunct{\mcitedefaultmidpunct}
	{\mcitedefaultendpunct}{\mcitedefaultseppunct}\relax
	\EndOfBibitem
	\bibitem[Bahabad \latin{et~al.}(2010)Bahabad, Murnane, and
	Kapteyn]{spatiotemporal}
	Bahabad,~A.; Murnane,~M.~M.; Kapteyn,~H.~C. Quasi-phase-matching of momentum
	and energy in nonlinear optical processes. \emph{Nat. Photonics}
	\textbf{2010}, \emph{4}, 570--575\relax
	\mciteBstWouldAddEndPuncttrue
	\mciteSetBstMidEndSepPunct{\mcitedefaultmidpunct}
	{\mcitedefaultendpunct}{\mcitedefaultseppunct}\relax
	\EndOfBibitem
	\bibitem[Bahabad \latin{et~al.}(2011)Bahabad, Murnane, and Kapteyn]{accelerate}
	Bahabad,~A.; Murnane,~M.~M.; Kapteyn,~H.~C. Manipulating nonlinear optical
	processes with accelerating light beams. \emph{Phys. Rev. A} \textbf{2011},
	\emph{84}, 033819\relax
	\mciteBstWouldAddEndPuncttrue
	\mciteSetBstMidEndSepPunct{\mcitedefaultmidpunct}
	{\mcitedefaultendpunct}{\mcitedefaultseppunct}\relax
	\EndOfBibitem
	\bibitem[Sivan and Pendry(2011)Sivan, and Pendry]{sivan2011theory}
	Sivan,~Y.; Pendry,~J.~B. Theory of wave-front reversal of short pulses in
	dynamically tuned zero-gap periodic systems. \emph{Phys. Rev. A}
	\textbf{2011}, \emph{84}, 033822\relax
	\mciteBstWouldAddEndPuncttrue
	\mciteSetBstMidEndSepPunct{\mcitedefaultmidpunct}
	{\mcitedefaultendpunct}{\mcitedefaultseppunct}\relax
	\EndOfBibitem
	\bibitem[Yaakobi and Friedland(2010)Yaakobi, and
	Friedland]{yaakobi2010autoresonant}
	Yaakobi,~O.; Friedland,~L. Autoresonant four-wave mixing in optical fibers.
	\emph{Phys. Rev. A} \textbf{2010}, \emph{82}, 023820\relax
	\mciteBstWouldAddEndPuncttrue
	\mciteSetBstMidEndSepPunct{\mcitedefaultmidpunct}
	{\mcitedefaultendpunct}{\mcitedefaultseppunct}\relax
	\EndOfBibitem
	\bibitem[Yaakobi \latin{et~al.}(2013)Yaakobi, Clerici, Caspani, Vidal, and
	Morandotti]{yaakobi2013complete}
	Yaakobi,~O.; Clerici,~M.; Caspani,~L.; Vidal,~F.; Morandotti,~R. Complete pump
	depletion by autoresonant second harmonic generation in a nonuniform medium.
	\emph{J. Opt. Soc. Am. B} \textbf{2013}, \emph{30}, 1637--1642\relax
	\mciteBstWouldAddEndPuncttrue
	\mciteSetBstMidEndSepPunct{\mcitedefaultmidpunct}
	{\mcitedefaultendpunct}{\mcitedefaultseppunct}\relax
	\EndOfBibitem
	\bibitem[Suchowski \latin{et~al.}(2014)Suchowski, Porat, and
	Arie]{suchowski2014adiabatic}
	Suchowski,~H.; Porat,~G.; Arie,~A. Adiabatic processes in frequency conversion.
	\emph{Laser Photonics Rev.} \textbf{2014}, \emph{8}, 333--367\relax
	\mciteBstWouldAddEndPuncttrue
	\mciteSetBstMidEndSepPunct{\mcitedefaultmidpunct}
	{\mcitedefaultendpunct}{\mcitedefaultseppunct}\relax
	\EndOfBibitem
	\bibitem[Moses \latin{et~al.}(2012)Moses, Suchowski, and
	K{\"a}rtner]{moses2012fully}
	Moses,~J.; Suchowski,~H.; K{\"a}rtner,~F.~X. Fully efficient adiabatic
	frequency conversion of broadband Ti: sapphire oscillator pulses. \emph{Opt.
		Lett.} \textbf{2012}, \emph{37}, 1589--1591\relax
	\mciteBstWouldAddEndPuncttrue
	\mciteSetBstMidEndSepPunct{\mcitedefaultmidpunct}
	{\mcitedefaultendpunct}{\mcitedefaultseppunct}\relax
	\EndOfBibitem
	\bibitem[Rangelov and Vitanov(2012)Rangelov, and
	Vitanov]{rangelov2012broadband}
	Rangelov,~A.~A.; Vitanov,~N.~V. Broadband sum-frequency generation using
	cascaded processes via chirped quasi-phase-matching. \emph{Phys. Rev. A}
	\textbf{2012}, \emph{85}, 045804\relax
	\mciteBstWouldAddEndPuncttrue
	\mciteSetBstMidEndSepPunct{\mcitedefaultmidpunct}
	{\mcitedefaultendpunct}{\mcitedefaultseppunct}\relax
	\EndOfBibitem
	\bibitem[Karenowska \latin{et~al.}(2012)Karenowska, Gregg, Tiberkevich, Slavin,
	Chumak, Serga, and Hillebrands]{karenowska2012oscillatory}
	Karenowska,~A.~D.; Gregg,~J.; Tiberkevich,~V.; Slavin,~A.; Chumak,~A.;
	Serga,~A.; Hillebrands,~B. Oscillatory energy exchange between waves coupled
	by a dynamic artificial crystal. \emph{Phys. Rev. Lett.} \textbf{2012},
	\emph{108}, 015505\relax
	\mciteBstWouldAddEndPuncttrue
	\mciteSetBstMidEndSepPunct{\mcitedefaultmidpunct}
	{\mcitedefaultendpunct}{\mcitedefaultseppunct}\relax
	\EndOfBibitem
	\bibitem[Sivan \latin{et~al.}(2016)Sivan, Rozenberg, Halstuch, and
	Ishaaya]{sivan2016nonlinear}
	Sivan,~Y.; Rozenberg,~S.; Halstuch,~A.; Ishaaya,~A. Nonlinear wave interactions
	between short pulses of different spatio-temporal extents. \emph{Sci. Rep.}
	\textbf{2016}, \emph{6}\relax
	\mciteBstWouldAddEndPuncttrue
	\mciteSetBstMidEndSepPunct{\mcitedefaultmidpunct}
	{\mcitedefaultendpunct}{\mcitedefaultseppunct}\relax
	\EndOfBibitem
	\bibitem[Boyd(2003)]{boyd}
	Boyd,~R.~W. \emph{Nonlinear optics}; Academic press, 2003\relax
	\mciteBstWouldAddEndPuncttrue
	\mciteSetBstMidEndSepPunct{\mcitedefaultmidpunct}
	{\mcitedefaultendpunct}{\mcitedefaultseppunct}\relax
	\EndOfBibitem
	\bibitem[Armstrong \latin{et~al.}(1962)Armstrong, Bloembergen, Ducuing, and
	Pershan]{armstrong1962interactions}
	Armstrong,~J.; Bloembergen,~N.; Ducuing,~J.; Pershan,~P. Interactions between
	light waves in a nonlinear dielectric. \emph{Phys. Rev} \textbf{1962},
	\emph{127}, 1918\relax
	\mciteBstWouldAddEndPuncttrue
	\mciteSetBstMidEndSepPunct{\mcitedefaultmidpunct}
	{\mcitedefaultendpunct}{\mcitedefaultseppunct}\relax
	\EndOfBibitem
	\bibitem[Zhang \latin{et~al.}(2007)Zhang, Lytle, Popmintchev, Zhou, Kapteyn,
	Murnane, and Cohen]{zhang2007quasi}
	Zhang,~X.; Lytle,~A.~L.; Popmintchev,~T.; Zhou,~X.; Kapteyn,~H.~C.;
	Murnane,~M.~M.; Cohen,~O. Quasi-phase-matching and quantum-path control of
	high-harmonic generation using counterpropagating light. \emph{Nat. Phys.}
	\textbf{2007}, \emph{3}, 270--275\relax
	\mciteBstWouldAddEndPuncttrue
	\mciteSetBstMidEndSepPunct{\mcitedefaultmidpunct}
	{\mcitedefaultendpunct}{\mcitedefaultseppunct}\relax
	\EndOfBibitem
	\bibitem[Akturk \latin{et~al.}(2010)Akturk, Gu, Bowlan, and
	Trebino]{akturk2010spatio}
	Akturk,~S.; Gu,~X.; Bowlan,~P.; Trebino,~R. Spatio-temporal couplings in
	ultrashort laser pulses. \emph{J. Opt.} \textbf{2010}, \emph{12},
	093001\relax
	\mciteBstWouldAddEndPuncttrue
	\mciteSetBstMidEndSepPunct{\mcitedefaultmidpunct}
	{\mcitedefaultendpunct}{\mcitedefaultseppunct}\relax
	\EndOfBibitem
	\bibitem[Konsens and Bahabad(2016)Konsens, and Bahabad]{konsens2016time}
	Konsens,~M.; Bahabad,~A. Time-to-frequency mapping of optical pulses using
	accelerating quasi-phase-matching. \emph{Phys. Rev. A} \textbf{2016},
	\emph{93}, 023823\relax
	\mciteBstWouldAddEndPuncttrue
	\mciteSetBstMidEndSepPunct{\mcitedefaultmidpunct}
	{\mcitedefaultendpunct}{\mcitedefaultseppunct}\relax
	\EndOfBibitem
	\bibitem[Bahabad \latin{et~al.}(2008)Bahabad, Cohen, Murnane, and
	Kapteyn]{bahabad2008quasi}
	Bahabad,~A.; Cohen,~O.; Murnane,~M.~M.; Kapteyn,~H.~C. Quasi-phase-matching and
	dispersion characterization of harmonic generation in the perturbative regime
	using counterpropagating beams. \emph{Opt. Express} \textbf{2008}, \emph{16},
	15923--15931\relax
	\mciteBstWouldAddEndPuncttrue
	\mciteSetBstMidEndSepPunct{\mcitedefaultmidpunct}
	{\mcitedefaultendpunct}{\mcitedefaultseppunct}\relax
	\EndOfBibitem
	\bibitem[Myer \latin{et~al.}(2014)Myer, Penfield, Gagnon, and Lytle]{Myer:14}
	Myer,~R.; Penfield,~A.; Gagnon,~E.; Lytle,~A.~L. Enhancing the Conversion
	Efficiency of Second Harmonic Generation Using Counterpropagating Light.
	Front. Opt. 2014. 2014; p FTh4C.4\relax
	\mciteBstWouldAddEndPuncttrue
	\mciteSetBstMidEndSepPunct{\mcitedefaultmidpunct}
	{\mcitedefaultendpunct}{\mcitedefaultseppunct}\relax
	\EndOfBibitem
	\bibitem[Eimerl \latin{et~al.}(1987)Eimerl, Davis, Velsko, Graham, and
	Zalkin]{eimerl1987optical}
	Eimerl,~D.; Davis,~L.; Velsko,~S.; Graham,~E.; Zalkin,~A. Optical, mechanical,
	and thermal properties of barium borate. \emph{J. Appl. Phys.} \textbf{1987},
	\emph{62}, 1968--1983\relax
	\mciteBstWouldAddEndPuncttrue
	\mciteSetBstMidEndSepPunct{\mcitedefaultmidpunct}
	{\mcitedefaultendpunct}{\mcitedefaultseppunct}\relax
	\EndOfBibitem
	\bibitem[Watanabe \latin{et~al.}(1993)Watanabe, Naito, and
	Chikama]{conjugation}
	Watanabe,~S.; Naito,~T.; Chikama,~T. Compensation of chromatic dispersion in a
	single-mode fiber by optical-phase conjugation. \emph{IEEE Photonics Technol.
		Lett.} \textbf{1993}, \emph{5}, 92--96\relax
	\mciteBstWouldAddEndPuncttrue
	\mciteSetBstMidEndSepPunct{\mcitedefaultmidpunct}
	{\mcitedefaultendpunct}{\mcitedefaultseppunct}\relax
	\EndOfBibitem
	\bibitem[Afanasjev \latin{et~al.}(1997)Afanasjev, Malomed, and Chu]{Chu}
	Afanasjev,~V.~V.; Malomed,~B.~A.; Chu,~P.~L. Dark soliton generation in a fused
	coupler. \emph{Opt. Commun.} \textbf{1997}, \emph{137}, 229--232\relax
	\mciteBstWouldAddEndPuncttrue
	\mciteSetBstMidEndSepPunct{\mcitedefaultmidpunct}
	{\mcitedefaultendpunct}{\mcitedefaultseppunct}\relax
	\EndOfBibitem
\end{mcitethebibliography}
\end{document}